\documentstyle[preprint,pre,aps]{revtex}

\tightenlines

\begin{document}

\draft

\title{Self-organizing height-arrow model:
numerical and analytical results}
\author{
Robert R.~Shcherbakov\thanks{On leave of absence from
Theoretical Department, Yerevan Physics Institute,
375036 Yerevan, Armenia.}
}
\address{Bogoliubov Laboratory of Theoretical Physics,
Joint Institute for Nuclear Research,
141980 Dubna, Russia.}
\date{\today}

\maketitle

\begin{abstract}
The recently introduced self-organizing height-arrow (HA) model
is numerically investigated on the square lattice and
analytically on the Bethe lattice. The concentration of occupied
sites and critical exponents of distributions of avalanches are
evaluated for two slightly different versions of the model.
The obtained exponents for distributions of avalanches by mass, area,
duration and appropriate fractal dimensions are close to those for
the BTW model, which suggests that the HA model belongs to the same
universality class. For comparison, the concentration of occupied
sites in the HA model is calculated exactly on the Bethe lattice
of coordination number $q=4$ as well.
\end{abstract}


\section{Introduction.}

\label{intro}

\setcounter{equation}0

The study of different cellular automata, which exhibit Self-Organized
Criticality (SOC)~\cite{BTW}, has been a subject of great interest in
recent years. These models serve as tractable limits of real dynamic
systems with many spatial degrees of freedom,
in which one might hope to gain understanding of
possible mechanisms of SOC. Unfortunately, in most cases
our current knowledge of the effects of SOC is rather limited.

The main peculiarity of the dynamic dissipative models
which are self-driven into the SOC state
is the presence of the power-law in distributions of
quantities such as avalanche mass, duration, etc.
These distributions are characterized by a set of exponents.
One of the most intriguing questions concerns the classes of
universality of these models. There are several attempts
to shed light on this problem~\cite{BB,NS}.
Recently, Ben-Hur and Biham~\cite{BB} proposed a classification
scheme for the $2d$ models both stochastic and deterministic.
They found that the original Bak, Tang, Wiesenfeld (BTW) model~\cite{BTW}
belongs to the universality class of {\it undirected} models,
{\it directed} models form a separate class,
and the two-state Manna model~\cite{Manna} belongs
to the universality class of {\it random relaxation} models.
Later on, Nakanishi and Sneppen~\cite{NS} examined several
$1d$ sandpile models and suggested that the two-state Manna model and
rice pile model~\cite{CCFFJ,PB} belong to the same universality class.

This classification scheme is based on the type of relaxations at
each site of the lattice. In the BTW model the particles from the toppled
sites are uniformly redistributed among its nearest neighbors,
whereas in the two-state Manna model the set of neighbors is chosen
randomly. It is possible to introduce more complicated dynamical
rules of relaxations at each site of the lattice that will depend on
some period $T$ where $T$ is the number of topplings.
During this period after each toppling the redistributions of particles
from the given site form a minimal nonperiodic sequence.
The BTW model can be considered with the period $T=1$. In the two-state
Mannna model the sequence of topplings at each site is stochastic
without any periodicity, therefore, one can put $T=\infty$ for this model.

In this paper, we investigate the recently introduced
self-organizing height-arrow (HA) model~\cite{PD,P2}.
It combines features of the BTW model~\cite{D,P1} and
self-organizing Eulerian walkers model (EWM)~\cite{PD,SPP}.
The model is a cellular automaton defined on any
connected undirected graph.
In this model, each site of the graph can be occupied by a particle or
can be empty. Addition of the particle to the occupied site
makes it unstable and causes its toppling. The site becomes empty
and the particles are transferred to the nearest neighbors.
The redistribution of particles from an unstable site is governed
by the second site variable, {\it an arrow}. Each outgoing particle
from the toppled site turns the arrow by the prescribed rule and
the new direction of the arrow determines the destination point
for this particle.

In the HA model $T$ is formed by the nonperiodic sequence of turns of
the arrow at each site of the lattice.
For simplicity, it is convenient to assign the same period $T$ for
all sites of the lattice. Thus, one can define the HA model with
increasingly complicated dynamical rules. These pseudo-random models
tend to the random ones for large $T \to \infty$.

The goal of the paper is the study of the HA model with $T=2$,
when two topplings restore initial direction of the arrow.
The model evolves at long times into the
steady state which is identified with the SOC state as
distributions of dynamic characteristics of the model
show a power-law behavior.
The obtained exponents for two slightly different types
of this model are very close to those for the BTW one, which suggests that
the HA model is in the universality class of {\it undirected} models.
We also study the main static characteristic of the model,
the time averaged density of occupied sites. This quantity
is obtained numerically on the square lattice and exactly
on the Bethe lattice of coordination number $q=4$.

The paper is organized as follows.
The definition of the model with two slightly different
types of evolution rules is presented in the next section.
Sec.~\ref{numeric} is devoted to the numerical investigation of the HA
model on the square lattice. Extrapolating results from finite size lattices
we estimate the concentration of occupied sites in the model.
The critical power-law exponents and scaling
relations among them are defined in the framework of finite-size
scaling analysis. We present results for the values of these exponents
in the SOC state for distributions of various quantities of the HA model.
Then, in Sec.~\ref{bethe}, we exactly calculate
the concentration of occupied sites on the Bethe lattice of
coordination number $q=4$.
Discussion and conclusions are presented in Sec.~\ref{conclus}.

\section{The self-organizing height-arrow model on the
square lattice.}

\label{model}

\setcounter{equation}0

The HA model we are going to investigate is defined as follows.
To each site $i$ of the two-dimensional $L\times L$ square lattice
is assigned a height variable $z_i\in \{0,1,...\}$
and an arrow directed north, east, south or west from $i$.
We start with an arbitrary initial configuration of heights and
arrows on the lattice. Initially, we drop a particle on the
randomly chosen site $i$.
The succeeding evolution of the system is determined by the following
rules. We increase the height variable at the site $i$ by 1,
$z_i\to z_i+1$. If the site $i$ is already occupied by the particle,
it topples ($z_i \to z_i-2$). To redistribute the particles
from the toppled site $i$ among its nearest neighbors, we turn
the arrow twice according to the prescribed rules.
For the given period $T=2$, there are only two non-equivalent sequences
of turns of the arrow at the given site
which preserve the model from being directed.
Hereafter, these sequences of turns will be
distinguished as {N-E-S-W-N} and {N-S-W-E-N}.
After each turn the
new direction of the arrow points to the neighbor sites to which
we will transfer particles at the next time step. This process
continues until a stable configuration is reached.
The sequence of topplings of unstable sites forms {\it an avalanche}
which propagates through the lattice.
After an avalanche ceases, we go on by adding a new
particle and so on.

A given configuration of the model is a set of directions of arrows
and heights. The total number of them is $8^{L\times L}$.
During the evolution of the system the arrow at any site might be only
in two positions due to the fact that the two subsequent toplings of
the site restore the initial position of the arrow.
Therefore, the set of configurations of the model falls into
$2^{L\times L}$ equivalent classes which are determined by
initial configurations of arrows.

Starting from a certain configuration of arrows and an arbitrary
configuration of occupied sites,
the model evolves through transient states into a dynamic attractor
which is critical.
This attractor is identified with the SOC state as different
dynamic characteristics of the
model show power-law tails in their distributions.
The model being in the SOC state passes from one allowed
stable configuration to another by avalanche dynamics.
This critical state has been investigated in detail by Priezzhev~\cite{P2}.
He defined operators
corresponding to addition of a particle at a randomly chosen site and
showed that they commute with each other. The algebra of these
operators is used to calculate the number of allowed
configurations of a given class in the SOC state. This number
is shown to be equal to the determinant of the discrete
Laplacian matrix $\Delta$ of the square lattice.
To check the given configuration to be
allowed in the SOC state the modification of the burning algorithm
was also introduced~\cite{P2}.

\section{Numerical results.}

\label{numeric}

\setcounter{equation}0

In order to investigate the static properties and avalanche dynamics
in the HA model, we have made numerical simulations with high
statistics. We consider square lattices of size $L\times L$ with
open boundary condition and $L$ ranging from $100$ up to $600$.
The HA model has been studied for two different types of dynamics
({N-E-S-W-N} and {N-S-W-E-N}) of turns of arrows and
various initial conditions.

Starting from an arbitrary distribution of occupied sites and certain
initial directions of arrows, the finite system evolves into a stationary
state.
In this state we have measured the time averaged density
$\langle p(z=1)\rangle$
and critical exponents for distributions of
avalanches by mass $(s)$, area $(a)$ and duration $(t)$.
The mass $s$ is defined as a total number of topplings in an avalanche
whereas the area $a$ is defined as a number of distinct sites
visited by an avalanche. The simultaneous topplings of
different sites in an avalanche at a given time is considered
a single time step. The duration $t$ is the number of this type of steps.
For a more detailed description of the structure of an avalanche
it is also useful to define a linear extent (diameter) of the
avalanche cluster via a radius of gyration ($r$).
We also measured the corresponding fractal dimensions
$\gamma_{xy}$, where $x,y=\{s,a,t,r\}$.

\begin{figure}[h]
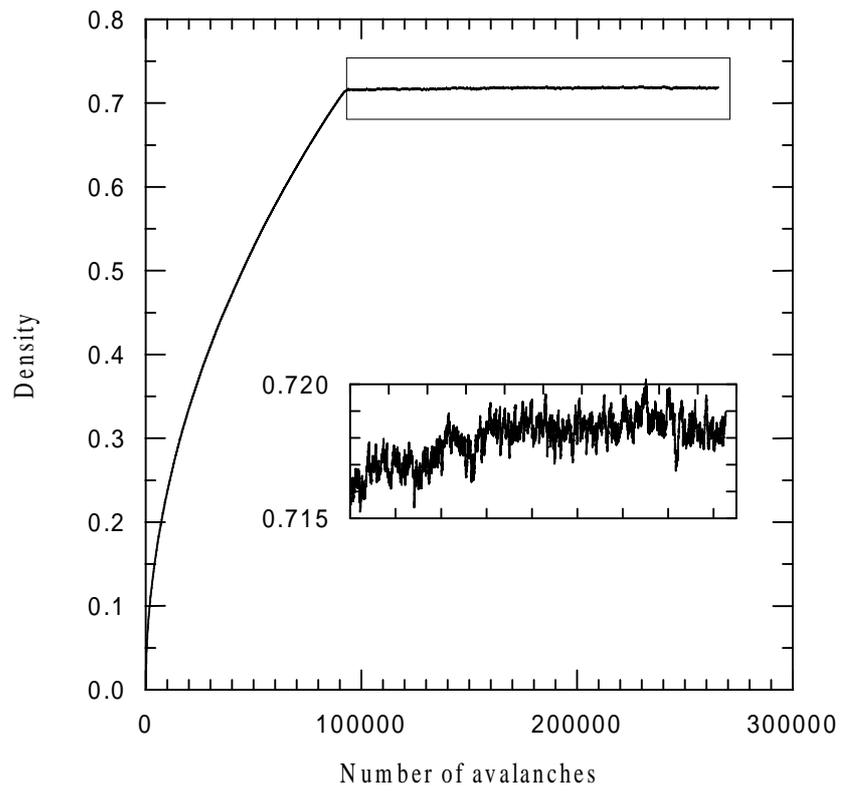

\unitlength 1mm
%%%%%%%%%%%%%%%%%%%%%%%%%%%%%%%%%%%%%%%%%%
%       Figure 3.1
%%%%%%%%%%%%%%%%%%%%%%%%%%%%%%%%%%%%%%%%%%
\caption{A computer simulation of the HA model ({N-E-S-W-N} dynamics)
on the square lattice of the linear size $L=600$ with open boundary
conditions.
The steady state is reached by the model after about 100 000 avalanches.}
\label{fig31}
\end{figure}

As is shown in Fig.~\ref{fig31}, the steady state is reached by the model
after about $100\,000$ avalanches on the square lattice of the linear
size $L=600$ for the system which is initially empty and with a random
initial distribution of arrows.
In this simulation, we were recording the averaged density
$\langle p(z=1)\rangle$ of occupied sites at each time when the
avalanche ceases.

As has been mentioned in Sec.~\ref{model}, the number of configurations
of the model falls into $2^{L\times L}$ classes depending on the
initial configurations of arrows. In our simulations of the HA model
with the {N-E-S-W-N} dynamics
we started from random initial configurations of arrows.
Whereas, for the {N-S-W-E-N} dynamics the arrows were initially
directed only east or south. The later case was chosen to simulate
the scattering of particles at each toppling by $180^0$ angle.

\begin{figure}[h]
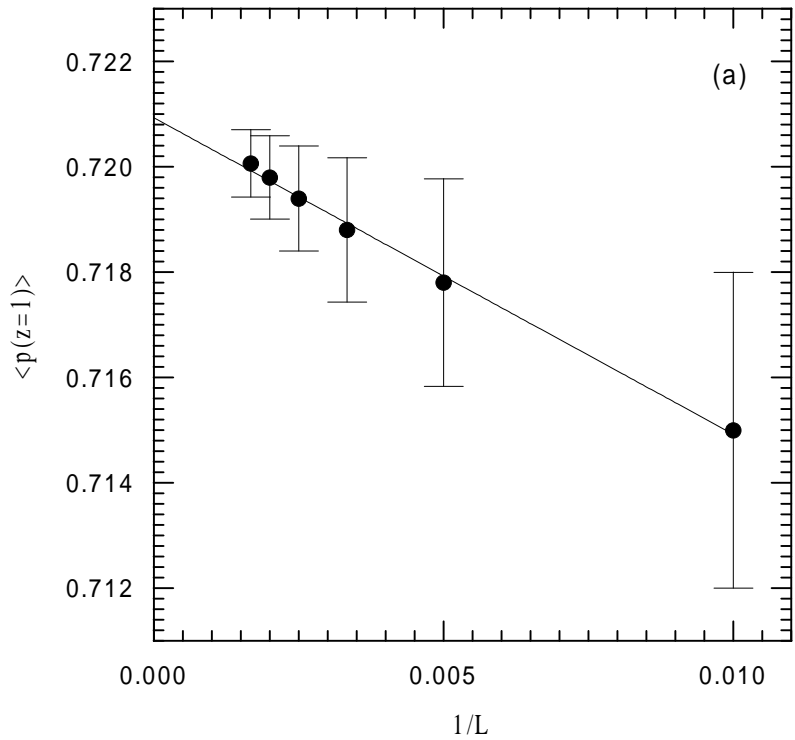

\unitlength 1mm
%%%%%%%%%%%%%%%%%%%%%%%%%%%%%%%%%%%%%%%%%%
%       Figure 3.2 a, b
%%%%%%%%%%%%%%%%%%%%%%%%%%%%%%%%%%%%%%%%%%
\caption{The dependence of the time averaged density of occupied
sites $\langle p(z=1)\rangle_L$ on the lattice size $L$.
(a) The HA model with {N-E-S-W-N} dynamics and random initial
directions of arrows.
(b) The same model with {N-S-W-E-N} dynamics and arrows initially
directed east or south. The numerical extrapolation
for the infinity lattice size $L\to \infty$ gives (a) $p_c=0.721\pm 0.001$
and (b) $p_c=0.755\pm 0.001$.}
\label{fig32}
\end{figure}

Fig.~\ref{fig32} displays the results of simulations for the time averaged density
$\langle p(z=1)\rangle$ in the stationary state.
They slightly depend on the lattice size $L$ and are well described
by the equation $\langle p(z=1)\rangle_L=p_c + c\,L^{-1}$.
The numerical extrapolation of the $L\to\infty$ limit gives the
values for the averaged density:
$p_c\equiv\lim_{L\to\infty}\langle p(z=1)\rangle_L=0.721 \pm 0.001$
({N-E-S-W-N} dynamics) and
$p_c=0.755 \pm 0.001$
({N-S-W-E-N} dynamics).
These values are a little higher in comparison with the
stochastic two-state Manna model~\cite{Manna} (see Table~\ref{table1}).

\begin{figure}[h]
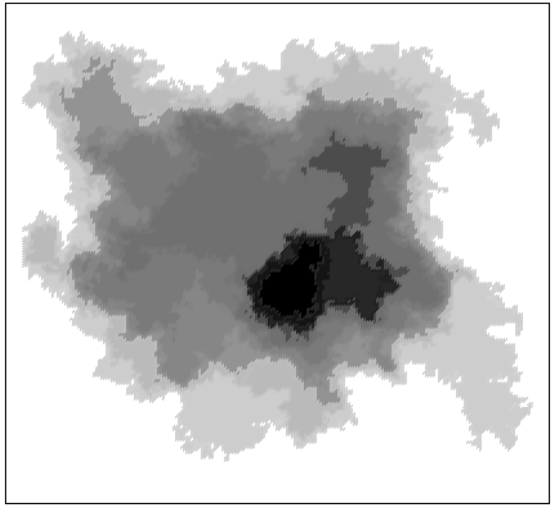

\unitlength 1mm
%%%%%%%%%%%%%%%%%%%%%%%%%%%%%%%%%%%%%%%%%%
%       Figure 3.3
%%%%%%%%%%%%%%%%%%%%%%%%%%%%%%%%%%%%%%%%%%
\caption{A typical form of an avalanche cluster of the HA model.
The lattice size is $L=200$. The avalanche cluster has a
layered structure. The number of topplings in each
layer is indicated in gray scale.}
\label{fig33}
\end{figure}

The form of avalanches in the HA model has a layered structure.
A typical avalanche is shown in Fig.~\ref{fig33} where the number of
relaxations in each site is marked by different scales of gray color.
The sites with
the same number of relaxations form {\it a layer} or {\it shell}.
We have observed that layers group in pairs.
In each pair a larger layer is a connected cluster with holes
whereas a smaller one is a disconnected cluster without holes.
Therefore, in the avalanche cluster there are a very few holes only
near the surface in the first layer where each site topples once.

\begin{figure}[h]
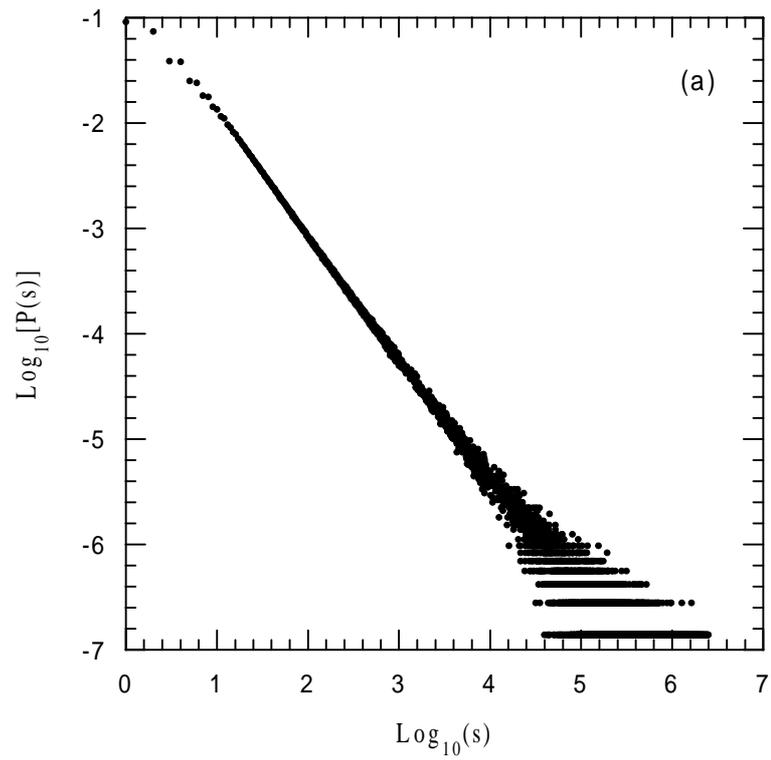

\unitlength 1mm
%%%%%%%%%%%%%%%%%%%%%%%%%%%%%%%%%%%%%%%%%%
%       Figure 3.4 a, b, c
%%%%%%%%%%%%%%%%%%%%%%%%%%%%%%%%%%%%%%%%%%
\caption{Simulation results for distributions of avalanches
by (a) mass (b) area and (c) duration of the HA model
in the SOC state. The linear size of the lattice is $L=600$.
The number of avalanches for each distribution is $10^7$.}
\label{fig34}
\end{figure}

In Figs.~\ref{fig34}, we present the directly measured distributions
of avalanches by mass $s$, area $a$ and duration $t$ in a
double logarithmic plot for the {N-E-S-W-N} dynamics
and the lattice of size $L=600$.
These distributions display a power-law behavior up to a certain
cutoff which depends on the system size $L$. Since our simulations
are limited by the lattices of finite size we ought to apply the finite-size
scaling analysis~\cite{KNWZ,B} assuming the distribution functions
scale with the lattice size $L$
\begin{equation}
\label{ansatz}
P(x,L)=L^{-\beta_x}f_x(x\cdot L^{-\nu_x})\,,
\end{equation}
where $f_x(xL^{-\nu_x})$ is a universal scaling function, $x$ stands for $s$,
$a$, $t$ or $r$, and $\beta_x$ and $\nu_x$ are critical exponents
which describe the scaling of the distribution function.
The finite-size scaling ansatz~(\ref{ansatz}) can be rewritten
in the following form~\cite{CO}:
\begin{equation}
\label{ansatz2}
P(x,L)=x^{-\beta_x/\nu_x}{\tilde f}_x(x\cdot L^{-\nu_x})\,.
\end{equation}

Let us suppose that distribution functions in the thermodynamic limit
$(L\to \infty)$ show pure power-law behavior for large enough
stochastic variables $(s,a,t,r)$
\begin{equation}
\label{distr}
P(x) \sim x^{-\tau_x}\,,\qquad x \gg 1\,,
\end{equation}
where $\tau_x$, $x\in \{s,a,t,r\}$ are critical exponents.
This conjecture is mainly supported by computer simulations
and different heuristic arguments\cite{CO}.
Therefore, comparing~(\ref{ansatz2}) and~(\ref{distr}) we get
{\it the scaling relations} among these exponents
\begin{equation}
\label{rel1}
\tau_x=\frac{\beta_x}{\nu_x}\,.
\end{equation}
From the fact that $\langle s \rangle \sim L^2$ in the undirected
BTW-type models~\cite{D}, one can get an additional scaling
relation~\cite{CO}
\begin{equation}
\label{rel11}
\nu_s(2-\tau_s)=2\,.
\end{equation}
If we also assume that the stochastic variables $s,a,t,r$ scale
against each other, the appropriate fractal dimensions
$\gamma_{xy}$ can be defined via the following relations~\cite{CFJ}:
\begin{equation}
\begin{array}{lll}
s \sim a^{\gamma_{sa}}\,, &\qquad & a \sim t^{\gamma_{at}}\,, \\
s \sim t^{\gamma_{st}}\,, &\qquad & a \sim r^{\gamma_{ar}}\,, \\
s \sim r^{\gamma_{sr}}\,, &\qquad & t \sim r^{\gamma_{tr}}\,,
\end{array}
\end{equation}
where $\gamma_{xy}=\gamma^{-1}_{yx}$.
The set of exponents $\{\tau_x\,,\gamma_{xy}\}$ are not independent
and scaling relations have the form~\cite{CFJ,CO}
\begin{equation}
\label{rel2}
\tau_x=1+\frac{\tau_y-1}{\gamma_{xy}}\,.
\end{equation}
From~(\ref{rel2}) one can find the simple expressions
\begin{equation}
\label{rel3}
\gamma_{xy}=\gamma_{xz}\gamma_{zy}\,.
\end{equation}
We have 10 unknown exponents altogether, namely $\tau_x$ and
$\gamma_{xy}=\gamma^{-1}_{yx}$, where $x,y\in\{a,s,t,r\}$,
but there exists only 6 linearly independent scaling
relations~(\ref{rel2}) among them.
Additional scaling relations can be obtained from the specific
structure and evolution of an avalanche and depend on the given
model.
The compactness of an avalanche cluster gives us
$\gamma_{ar}=2$~\cite{MD,CO}.

Thus, estimating only three critical exponents from the numerical data,
we can calculate all the others using the scaling relations
Eqs.~(\ref{rel2}).
Having calculated more than three exponents we are able to
check these relations as well.
The accurate determination of the $\tau$'s exponents is a more difficult
task than the $\gamma$'s due to the strict dependence of the
$\tau$'s on the system size $L$.

\begin{figure}[h]
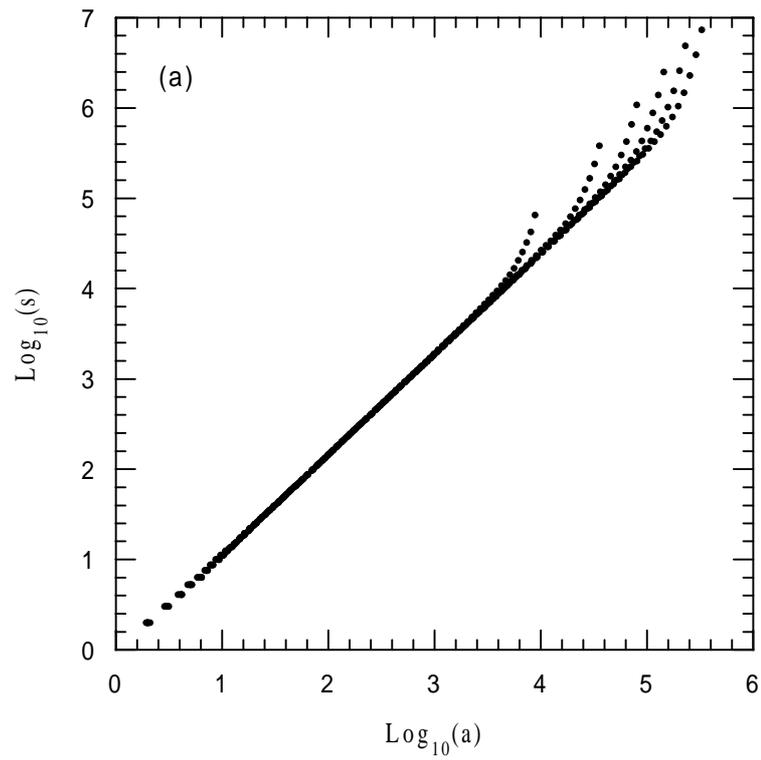

\unitlength 1mm
%%%%%%%%%%%%%%%%%%%%%%%%%%%%%%%%%%%%%%%%%%
%       Figure 3.5 a, b, c
%%%%%%%%%%%%%%%%%%%%%%%%%%%%%%%%%%%%%%%%%%
\caption{Double-logarithmic plot of the dependence
of the stochastic variables $\{s, a, t\}$ against each other for different
lattice sizes.
The distributions are integrated over bin lengths.}
\label{fig37}
\end{figure}

To reduce the fluctuations of the data, we integrated each
distribution over bin lengths.
The exponents $\gamma_{xy}$, $x,y=\{s,a,t\}$, are measured from the slopes
of the straight parts of the corresponding graphs (Figs.~\ref{fig37}).
The obtained values are shown in Table~\ref{table2}.

\begin{figure}[h]
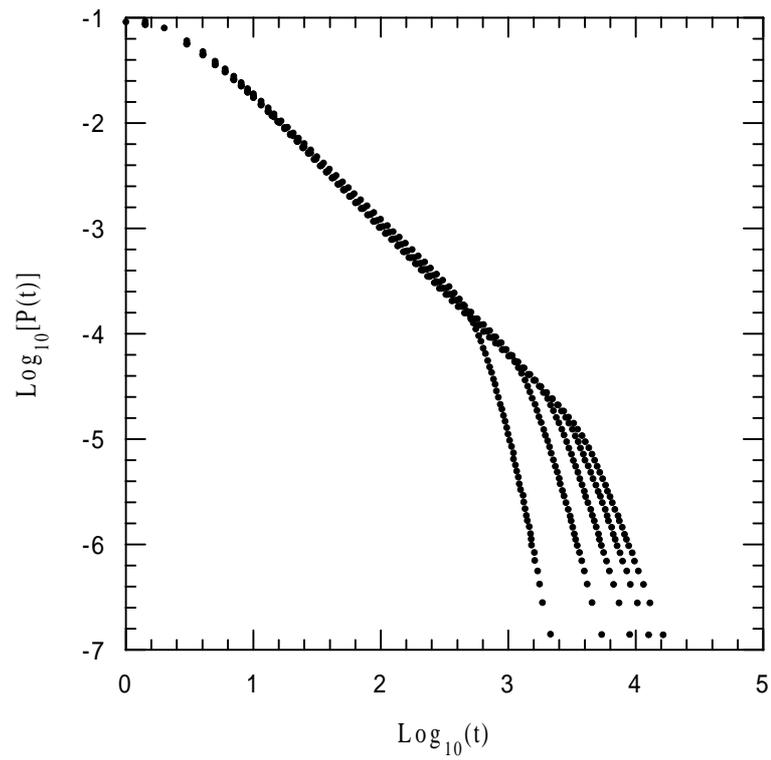

\unitlength 1mm
%%%%%%%%%%%%%%%%%%%%%%%%%%%%%%%%%%%%%%%%%%
%       Figure 3.6
%%%%%%%%%%%%%%%%%%%%%%%%%%%%%%%%%%%%%%%%%%
\caption{Double-logarithmic plot of the integrated
distribution of avalanches $P(t,L)$ versus duration $t$ for
five lattice sizes $L=100,200,...,500$. Each distribution
is averaged over $10^{7}$ avalanches.}
\label{fig310}
\end{figure}

\begin{figure}[h]
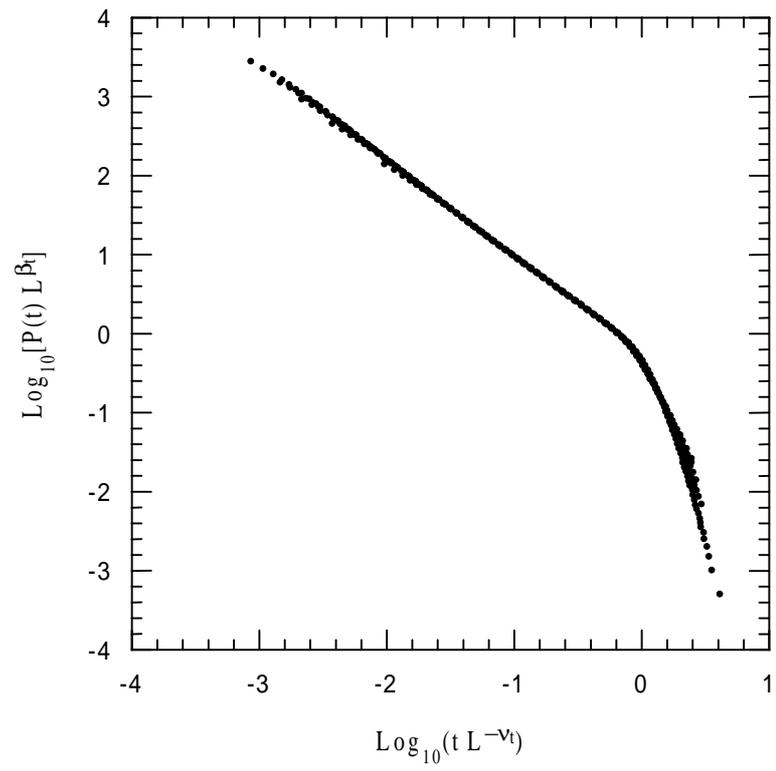

\unitlength 1mm
%%%%%%%%%%%%%%%%%%%%%%%%%%%%%%%%%%%%%%%%%%
%       Figure 3.7
%%%%%%%%%%%%%%%%%%%%%%%%%%%%%%%%%%%%%%%%%%
\caption{Finite-size scaling plot for the integrated distributions $P(t,L)$.
The data for different $L$ collapse onto a single curve for $\beta_t=1.78$
and $\nu_t=1.36$}
\label{fig311}
\end{figure}

Plotting integrated distributions $P(t,L)\cdot L^{\beta_t}$ versus
$t\cdot L^{-\nu_t}$ on a double
logarithmic scale, as is shown in Fig.~\ref{fig310} for different
lattice sizes $L$, we obtained from finite-size scaling analysis
that the best data collapse corresponds to
$\beta_t=1.78\pm 0.05$, $\nu_t=1.36\pm 0.05$ (Fig.~\ref{fig311}).
The scaling relation for the critical exponents~(\ref{rel1})
gives the value $\tau_t=1.31\pm 0.05$.

Next, we use the measured values of $\tau_t$, $\gamma_{st}$ and
$\gamma_{sa}$ to estimate the whole set of exponents using
the scaling relations, Eqs.~(\ref{rel2}).
These values are presented in Table~\ref{table2}.

\begin{table*}[h]
\caption{The time averaged density $p_c$ of occupied sites for the HA model
on the square lattice with two slightly different types of dynamics
and on the Bethe lattice is compared with the value for the two-state
Manna model.
The uncertainty of the numerical data is about $\pm 0.001$.}
\label{table1}
\begin{tabular}{lllll}
        &     \multicolumn{4}{c}{Model}  \\ \cline{2-5}
Density &HA\tablenote{{N-E-S-W-N} dynamics} &
         HA\tablenote{{N-S-W-E-N} dynamics} &
         HA\tablenote{Bethe lattice} & Manna~\cite{Manna}\\ \hline
$p_c$   &$0.721$&$0.755$&$\frac{2}{3}\approx 0.666$
        &$0.683$  \\
\end{tabular}
\end{table*}

\begin{table*}[h]
\caption{The critical exponents for the $2d$ HA model evaluated in our
work (first column) are compared with those for the BTW
and two-state Manna models.
The second column is the critical exponents for the BTW model
obtained from numerical simulations, whereas in the third column
we show exact values of the exponents for the BTW model based on the
scaling relations~(\ref{rel2}), (\ref{rel3}) and
$\gamma_{sr}=\tau_r +1$~[16].
Comparison of the critical exponents of the HA and BTW models
evaluated from numerical simulations shows that the HA model belongs
to the universality class of the BTW model.
The uncertainty of the numerical
data for the HA model is about $\pm 0.05$.}
\label{table2}
\begin{tabular}{lllll}
              & \multicolumn{4}{c}{Model} \\ \cline{2-5}
Exponent&HA&BTW & BTW\tablenote{Exact result.}&Manna\\ \hline
$\tau_s$      & $1.18$\tablenote{The value of the exponent is obtained
from the scaling relations~(\ref{rel2}) and~(\ref{rel3}).}
              &1.20~\cite{Manna2} & $\frac{6}{5}=1.2$~\cite{PKI} & $1.30$~\cite{Manna}   \\
$\tau_a$      &$1.21^{\rm b}$ & $1.22$~\cite{Manna3}& $\frac{5}{4}=1.25$~\cite{PKI}& $1.37^{\rm b}$ \\
$\tau_t$      &$1.31$         & 1.32~\cite{Manna2} & $\frac{7}{5}=1.4^{\rm b}$    & $1.50$~\cite{Manna} \\
$\tau_r$      &$1.41^{\rm b}$ & $1.42^{\rm b}$     & $\frac{3}{2}=1.5^{\rm b}$    & $1.75^{\rm b}$   \\
$\gamma_{sa}$ &$1.11$         & 1.06~\cite{BB}     & $\frac{5}{4}=1.25^{\rm b}$   & $1.23$~\cite{BB} \\
$\gamma_{st}$ &$1.68$         & 1.64~\cite{Manna2} & $2^{\rm b}$                  & $1.67$~\cite{Manna}   \\
$\gamma_{sr}$ &$2.23^{\rm b}$ & $2.16^{\rm b}$     & $\frac{5}{2}=2.5^{\rm b}$    & $2.49^{\rm b}$ \\
$\gamma_{at}$ &$1.51$         & $1.52^{\rm b}$     & $\frac{8}{5}=1.6^{\rm b}$    & $1.35$~\cite{BB} \\
$\gamma_{ar}$ &$2$\tablenote{From the compactness of an avalanche cluster~\cite{MD}.} &
              $2^{\rm c}$      & $2^{\rm c}$                & $2.01^{\rm b}$  \\
$\gamma_{tr}$ &$1.33^{\rm b}$ & 1.32~\cite{BB}     & $\frac{5}{4}=1.25$~\cite{MD} & $1.49$~\cite{BB} \\
\end{tabular}
\end{table*}

The simulations for the HA model with {N-S-W-E-N} dynamics
within a small uncertainty give
the same values for the critical exponents.

\section{The height-arrow model on the Bethe lattice.}

\label{bethe}

\setcounter{equation}0

In this section, we present exact analytical calculations
for the averaged density of occupied sites in the HA model on
the Bethe lattice of coordination number $q=4$.
The Bethe lattice is defined through {\it a Cayle tree} well-known
in graph theory which is a connected graph with no closed
circuits of edges. Then, the Bethe lattice is an infinite Cayle
tree homogenous in the sense that all except the outer vertices
have the same coordination number $q$~\cite{Baxter}.

Following Dhar~\cite{DM}, we approach the problem by
dividing the allowed configurations of the HA model in the SOC state
into two types: {\it strongly allowed} and {\it weakly allowed}
and constructing the recurrent relations for the
ratio of these configurations on the branches of the Bethe lattice.
Using this ratio in the thermodynamic limit, we obtain the density
of occupied sites in the HA model.

\begin{figure}[h]
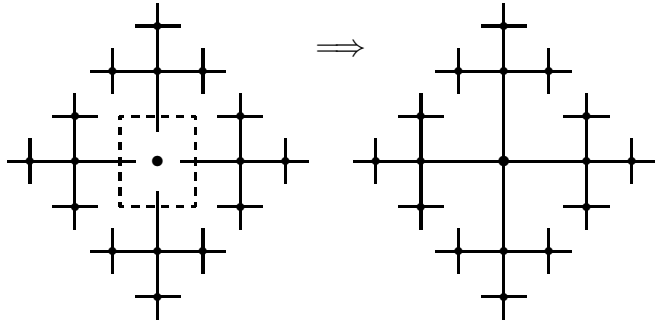

\unitlength 1mm
\thicklines
%%%%%%%%%%%%%%%%%%%%%%%%%%%%%%%%%%%%%%%%%%
%       Figure 4.1
%%%%%%%%%%%%%%%%%%%%%%%%%%%%%%%%%%%%%%%%%%
\caption{Construction of the Cayle tree with $q=4$ and $k=3$
generations by attaching $q=4$ $k$th-generation branches to
a central site. This procedure is explained in the text.}
\label{fig41}
\end{figure}

First, let us briefly describe the procedure of construction of
the Cayle tree. Like many tree-like structures, the Cayle tree
of $k$ generations of coordination number $q$
can be constructed by attaching $q$ $k$th-generation
branches to a central site, as is shown in Fig.~\ref{fig41}.
In turn, every $k$th-generation branch
is constructed by connecting $q-1$ $(k-1)$th-generation
branches to a new root and so on~\cite{Baxter}.
This property allows us to build recursion relations for the number
of allowed configurations on the branches of the Cayle tree.

The number of boundary sites of the Cayle tree is comparable with
interior ones. Hence, the calculation of the bulk properties in the
thermodynamic limit requires special care. Since we are interested
in the solution on the Bethe lattice, we will take the result for
the averaged density of occupied sites calculated at the central
site of the Cayle tree as the value for the Bethe lattice.

The definition of the HA model on this connected graph of coordination
number $q=4$ remains unchanged.
The only difference from the square lattice concerns the
notation of directions of arrows and sequences of their turns.
We will consider only sequential clockwise turns by the right angle
and denote the directions of arrows at each site simply by $\{1,2,3,4\}$.

\begin{figure}[h]
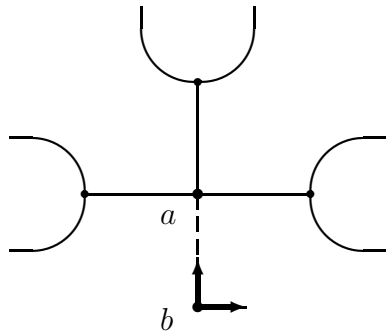

\unitlength 1mm
\thicklines
%%%%%%%%%%%%%%%%%%%%%%%%%%%%%%%%%%%%%%%%%%
%       Figure 4.2
%%%%%%%%%%%%%%%%%%%%%%%%%%%%%%%%%%%%%%%%%%
\caption{A $k$th-generation branch $T_k$ and vertex $b$ form a
subgraph $T'$. The ovals denote the rest of the subbranches of $T_k$.}
\label{fig42}
\end{figure}

Let $C$ be an allowed configuration on the $k$th-generation branch
$T_k$ with root vertex $a$. Adding a vertex $b$ to $T_k$, one
defines a subgraph $T'=T_k\cup b$. If the subconfiguration $C'$
on $T'$ with $z_b=0$ and an arrow directed up or right
(Fig.~\ref{fig42})
becomes forbidden, $C$ is called {\it a weakly allowed} (W)
configuration, otherwise it is called {\it a strongly allowed} (S) one.

\begin{figure}[h]
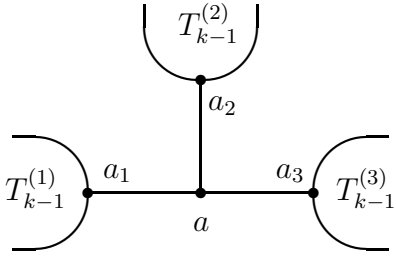

\unitlength 1mm
\thicklines
%%%%%%%%%%%%%%%%%%%%%%%%%%%%%%%%%%%%%%%%%%
%       Figure 4.3
%%%%%%%%%%%%%%%%%%%%%%%%%%%%%%%%%%%%%%%%%%
\caption{A $k$th-generation branch $T_k$ consists of three
nearest $(k-1)$th-generation branches $T^{(1)}_{k-1}$,
$T^{(2)}_{k-1}$ and $T^{(3)}_{k-1}$.}
\label{fig43}
\end{figure}

Now consider $T_k$ with a root $a$ that consists of three
$(k-1)$th-generation branches $T^{(1)}_{k-1}$, $T^{(2)}_{k-1}$ and
$T^{(3)}_{k-1}$ with roots $a_1$, $a_2$ and $a_3$, respectively
(Fig.~\ref{fig43}). Let $N_{W}(T_k,n,\uparrow)$
and $N_{S}(T_k,n,\uparrow)$
be the numbers of distinct $W$- and $S$-type configurations
on $T_k$ with a given height $z_a=n$ and direction
of the arrow at the root vertex $a$.

Let us also introduce
\begin{equation}
\label{f4.1}
N_{W}(T_k)=\sum\limits_{n=\{0,1\}}
\sum\limits_{r=\{\uparrow,\downarrow \}}
N_{W}(T_k,n,r)\,,
\end{equation}
\begin{equation}
\label{f4.2}
N_{S}(T_k)=\sum\limits_{n=\{0,1\}}
\sum\limits_{r=\{\uparrow,\downarrow \}}
N_{S}(T_k,n,r)\,,
\end{equation}
where the first summation is over the values of the heights and the second
one is over the directions of the arrow. As has been already mentioned,
the arrow at each site can take only two directions.

These numbers can be expressed in terms of the numbers of allowed
subconfigurations on the three $(k-1)$th-generation branches
$T^{(1)}_{k-1}$, $T^{(2)}_{k-1}$ and $T^{(3)}_{k-1}$:
\begin{eqnarray}
\label{f4.3}
N_W(T_k) & = & N_S^{(1)}N_S^{(2)}N_S^{(3)} + N_S^{(1)}N_S^{(2)}N_W^{(3)}+
               N_S^{(1)}N_W^{(2)}N_S^{(3)} + N_S^{(1)}N_W^{(2)}N_W^{(3)}+
               \nonumber\\
         &   & N_W^{(1)}N_S^{(2)}N_S^{(3)} + N_W^{(1)}N_S^{(2)}N_W^{(3)}+
               N_W^{(1)}N_W^{(2)}N_S^{(3)} + N_W^{(1)}N_W^{(2)}N_W^{(3)}\,,\\
\label{f4.4}
N_S(T_k) & = & 3N_S^{(1)}N_S^{(2)}N_S^{(3)} + 2N_S^{(1)}N_S^{(2)}N_W^{(3)}+
	       2N_S^{(1)}N_W^{(2)}N_S^{(3)} + N_S^{(1)}N_W^{(2)}N_W^{(3)}+
	       \nonumber\\
	 &   & 2N_W^{(1)}N_S^{(2)}N_S^{(3)} + N_W^{(1)}N_S^{(2)}N_W^{(3)}+
	       N_W^{(1)}N_W^{(2)}N_S^{(3)}\,,
\end{eqnarray}
where $N_{\alpha}^{(i)}\equiv N_{\alpha}(T^{(i)}_{k-1})$, $\alpha=W,S$
and $i=1,2,3$.

Let us define
\begin{equation}
\label{f4.5}
X=\frac{N_W}{N_S}\,.
\end{equation}

If we consider graphs $T^{(1)}_{k-1}$, $T^{(2)}_{k-1}$ and $T^{(3)}_{k-1}$
to be isomorphic, then
$N(T^{(1)}_{k-1})=N(T^{(2)}_{k-1})=N(T^{(3)}_{k-1})$ and
from~(\ref{f4.3}) and (\ref{f4.4}) one obtains the following
recursion relation:
\begin{equation}
\label{f4.7}
X(T_k)= \frac{1}{3}\left(1+X(T_{k-1})\right)\,.
\end{equation}
With the initial condition $X(T_0)=\frac{1}{3}$, this equation has a simple
solution
\begin{equation}
\label{f4.8}
X(T_k)= \frac{1}{2} - \frac{1}{2}\,3^{-(k+1)}\,.
\end{equation}
In the thermodynamic limit $(k\to\infty)$ the iterative
sequence $\{X(T_k)\}$ converges to a stable point
$X^*=\frac{1}{2}$ that characterizes the ratio of the weakly allowed
configurations to the strongly allowed ones in the SOC state.

\begin{figure}[h]
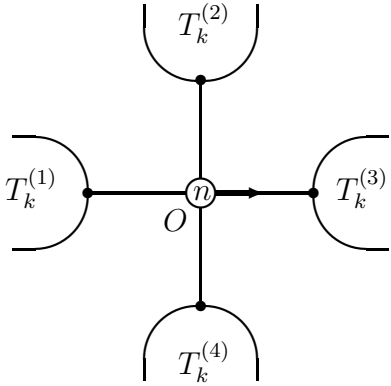

\unitlength 1mm
\thicklines
%%%%%%%%%%%%%%%%%%%%%%%%%%%%%%%%%%%%%%%%%%
%       Figure 4.4
%%%%%%%%%%%%%%%%%%%%%%%%%%%%%%%%%%%%%%%%%%
\caption{
A site $O$ with height $z_o=n$ and a given direction of the arrow
is located deep inside the lattice
and surrounded by the four $k$th-generation branches
$T_k^{(1)}, T_k^{(2)}, T_k^{(3)}$ and $T_k^{(4)}$.}
\label{fig44}
\end{figure}

Consider now a randomly chosen site $O$ deep inside the
Cayle tree (Fig.~\ref{fig44}).
The probability $P(1)$ of occupation of the site $O$ is
\begin{equation}
\label{f4.9}
P(1)=\frac{N(1)}{N_{\mbox{\scriptsize total}}}\,,
\end{equation}
where $N(1)$ is the number of allowed configurations with
$z=1$ at the site $O$ and
$N_{\mbox{\scriptsize total}}=N(0)+N(1)$ is
the total number of allowed configurations on the Cayle tree.
The numbers $N(0)$ and $N(1)$ can be expressed via
the numbers of allowed configurations on the four neighbor
$k$th-generation branches $T^{(i)}_k$, $i=1,2,3,4$
\begin{eqnarray}
\label{f4.10}
N(0) & = & 2[1+2X+X^2]\prod\limits_{i=1}^{4}N_S(T^{(i)})\,,\\
\label{f4.11}
N(1) & = & 2[1+4X+5X^2+2X^3]\prod\limits_{i=1}^{4}N_S(T^{(i)})\,.
\end{eqnarray}
For the sites far from the surface in the thermodynamic limit
$(k\rightarrow\infty)$ we have $X=\frac{1}{2}$.
Thus, from(~\ref{f4.10}) and (\ref{f4.11}) we obtain
\begin{equation}
\label{f4.12}
P(0)=\frac{1}{3}\,,\qquad P(1)=\frac{2}{3}\,.
\end{equation}
The value for the concentration of occupied sites $P(1)$
is in good qualitative agreement with the numerical result obtained
on the square lattice.

\section{Conclusion.}

\label{conclus}

We numerically studied the self-organizing height-arrow (HA) model
on the square lattice and analytically on the Bethe lattice.
The dynamics of the model drives it into the critical attractor with
spatio-temporal complexity. The obtained distributions of various
dynamic characteristics show an explicit power law behavior which
indicates long-range correlations in the steady state of the system.
To obtain the critical exponents of distributions of dynamic quantities
of the model in the SOC state, we applied the finite-size
scaling analysis. The values of these exponents are listed in
Table~\ref{table2} and compared with known exponents of the BTW model
and two-state Manna model. Thus, we argue that the HA model belongs
to the universality class of undirected models.

Furthermore, we investigated the averaged density of occupied sites $p_c$ in
the SOC state of the HA model. It was also calculated exactly on the
Bethe lattice of coordination number $q=4$. The obtained results are
presented in Table~\ref{table1}.

\vspace{2cm}

\section*{Acknowledgments}

I would like to thank V.B.~Priezzhev for valuable contributions
and suggesting improvements.
I thank N.S.~Ananikian, E.V.~Ivashkevich, D.V.~Ktitarev, Vl.V.~Papoyan,
A.M.~Povolotsky and B.~Tadic for fruitful discussions.
Partial financial support by the Russian Foundation for Basic Research
under grant No. 97-01-01030 is acknowledged.

%\newpage

%\onecolumn

\newpage

\pagestyle{empty}

\setcounter{figure}{7}

\begin{figure}[h]
\unitlength 1mm
\thicklines
%%%%%%%%%%%%%%%%%%%%%%%%%%%%%%%%%%%%%%%%%%
%       Figure 4.1
%%%%%%%%%%%%%%%%%%%%%%%%%%%%%%%%%%%%%%%%%%
\begin{center}
\begin{picture}(86,45)
%
% 4 kth-branches
%
\put(20,21){\circle*{1.5}}
\put(15,15){\dashbox{1}(10,12){\ }}
\put(0,21){\line(1,0){17}}
\put(9,12){\line(0,1){18}}
\put(6,15){\line(1,0){6}}
\put(3,18){\line(0,1){6}}
\put(6,27){\line(1,0){6}}
\put(9,21){\circle*{1}}
\put(9,15){\circle*{1}}
\put(3,21){\circle*{1}}
\put(9,27){\circle*{1}}
\put(20,25){\line(0,1){17}}
\put(11,33){\line(1,0){18}}
\put(14,30){\line(0,1){6}}
\put(17,39){\line(1,0){6}}
\put(26,30){\line(0,1){6}}
\put(20,33){\circle*{1}}
\put(14,33){\circle*{1}}
\put(20,39){\circle*{1}}
\put(26,33){\circle*{1}}
\put(23,21){\line(1,0){17}}
\put(31,12){\line(0,1){18}}
\put(28,27){\line(1,0){6}}
\put(37,18){\line(0,1){6}}
\put(28,15){\line(1,0){6}}
\put(31,21){\circle*{1}}
\put(31,27){\circle*{1}}
\put(37,21){\circle*{1}}
\put(31,15){\circle*{1}}
\put(20,0){\line(0,1){17}}
\put(11,9){\line(1,0){18}}
\put(14,6){\line(0,1){6}}
\put(17,3){\line(1,0){6}}
\put(26,6){\line(0,1){6}}
\put(20,9){\circle*{1}}
\put(14,9){\circle*{1}}
\put(26,9){\circle*{1}}
\put(20,3){\circle*{1}}
\put(41,35){$\Longrightarrow$}
\put(66,21){\circle*{1.5}}
\put(46,21){\line(1,0){21}}
\put(55,12){\line(0,1){18}}
\put(52,15){\line(1,0){6}}
\put(49,18){\line(0,1){6}}
\put(52,27){\line(1,0){6}}
\put(55,21){\circle*{1}}
\put(55,15){\circle*{1}}
\put(49,21){\circle*{1}}
\put(55,27){\circle*{1}}
\put(66,21){\line(0,1){21}}
\put(57,33){\line(1,0){18}}
\put(60,30){\line(0,1){6}}
\put(63,39){\line(1,0){6}}
\put(72,30){\line(0,1){6}}
\put(66,33){\circle*{1}}
\put(60,33){\circle*{1}}
\put(66,39){\circle*{1}}
\put(72,33){\circle*{1}}
\put(65,21){\line(1,0){21}}
\put(77,12){\line(0,1){18}}
\put(74,27){\line(1,0){6}}
\put(83,18){\line(0,1){6}}
\put(74,15){\line(1,0){6}}
\put(77,21){\circle*{1}}
\put(77,27){\circle*{1}}
\put(83,21){\circle*{1}}
\put(77,15){\circle*{1}}
\put(66,0){\line(0,1){21}}
\put(57,9){\line(1,0){18}}
\put(60,6){\line(0,1){6}}
\put(63,3){\line(1,0){6}}
\put(72,6){\line(0,1){6}}
\put(66,9){\circle*{1}}
\put(60,9){\circle*{1}}
\put(72,9){\circle*{1}}
\put(66,3){\circle*{1}}
\end{picture}
\end{center}
\caption{Construction of the Cayle tree with $q=4$ and $k=3$
generations by attaching $q=4$ $k$th-generation branches to
a central site. This procedure is explained in the text.}
\end{figure}

\begin{figure}[h]
\unitlength 1mm
\thicklines
%%%%%%%%%%%%%%%%%%%%%%%%%%%%%%%%%%%%%%%%%%
%       Figure 4.2
%%%%%%%%%%%%%%%%%%%%%%%%%%%%%%%%%%%%%%%%%%
\begin{center}
\begin{picture}(70,55)
\put(35,25){\line(0,1){15}}
\put(20,25){\line(1,0){30}}
%
% I-sector
%
\put(10,25){\oval(20,15)[r]}
%
% II-sector
%
\put(35,50){\oval(15,20)[b]}
%
% III-sector
%
\put(60,25){\oval(20,15)[l]}
\put(20,25){\circle*{1}}
\put(35,40){\circle*{1}}
\put(50,25){\circle*{1}}
\put(35,25){\circle*{1.5}}
\put(30,21){$a$}
\multiput(35,25)(0,-3){3}{\line(0,-1){2}}
\put(35,10){\circle*{1.5}}
\put(30,7){$b$}
\put(35,10){\vector(0,1){6.5}}
\put(34.8,10){\line(0,1){5}}
\put(35.2,10){\line(0,1){5}}
\put(35,10){\vector(1,0){6.5}}
\put(35,9.8){\line(1,0){5}}
\put(35,10.2){\line(1,0){5}}
\end{picture}
\end{center}
\caption{A $k$th-generation branch $T_k$ and vertex $b$ form a
subgraph $T'$. The ovals denote the rest of the subbranches of $T_k$.}
\end{figure}

\newpage

\begin{figure}[h]
\unitlength 1mm
\thicklines
%%%%%%%%%%%%%%%%%%%%%%%%%%%%%%%%%%%%%%%%%%
%       Figure 4.3
%%%%%%%%%%%%%%%%%%%%%%%%%%%%%%%%%%%%%%%%%%
\begin{center}
\begin{picture}(70,40)
\put(35,15){\line(0,1){15}}
\put(20,15){\line(1,0){30}}
%
% I-sector
%
\put(10,15){\oval(20,15)[r]}
\put(9,14){$T^{(1)}_{k-1}$}
\put(22,17){$a_1$}
%
% II-sector
%
\put(35,40){\oval(15,20)[b]}
\put(32,36){$T^{(2)}_{k-1}$}
\put(36,26){$a_2$}
%
% III-sector
%
\put(60,15){\oval(20,15)[l]}
\put(53,14){$T^{(3)}_{k-1}$}
\put(45,17){$a_3$}
\put(20,15){\circle*{1.5}}
\put(35,30){\circle*{1.5}}
\put(50,15){\circle*{1.5}}
\put(35,15){\circle*{1.5}}
\put(34,10){$a$}
\end{picture}
\end{center}
\caption{A $k$th-generation branch $T_k$ consists of three
nearest $(k-1)$th-generation branches $T^{(1)}_{k-1}$,
$T^{(2)}_{k-1}$ and $T^{(3)}_{k-1}$.}
\end{figure}

\begin{figure}[h]
\unitlength 1mm
\thicklines
%%%%%%%%%%%%%%%%%%%%%%%%%%%%%%%%%%%%%%%%%%
%       Figure 4.4
%%%%%%%%%%%%%%%%%%%%%%%%%%%%%%%%%%%%%%%%%%
\begin{center}
\begin{picture}(70,70)
\put(35,20){\line(0,1){13}}
\put(35,50){\line(0,-1){13}}
\put(20,35){\line(1,0){13}}
\put(50,35){\line(-1,0){13}}
%
% I-sector
%
\put(10,35){\oval(20,15)[r]}
\put(9,34){$T^{(1)}_{k}$}
%
% II-sector
%
\put(35,60){\oval(15,20)[b]}
\put(32,56){$T^{(2)}_{k}$}
%
% III-sector
%
\put(60,35){\oval(20,15)[l]}
\put(53,34){$T^{(3)}_{k}$}
%
% IV-sector
%
\put(35,10){\oval(15,20)[t]}
\put(32,11){$T^{(4)}_{k}$}
\put(20,35){\circle*{1.5}}
\put(35,50){\circle*{1.5}}
\put(50,35){\circle*{1.5}}
\put(35,20){\circle*{1.5}}
\put(35,35){\circle{4}}
\put(34,34){$n$}
\put(30,30){$O$}
\put(37,35){\vector(1,0){6.5}}
\put(37,34.8){\line(1,0){5}}
\put(37,35.2){\line(1,0){5}}
\end{picture}
\end{center}
\caption{A site $O$ with height $z_o=n$ and a given direction of the arrow
is located deep inside the lattice
and surrounded by the four $k$th-generation branches
$T_k^{(1)}, T_k^{(2)}, T_k^{(3)}$ and $T_k^{(4)}$.}
\end{figure}

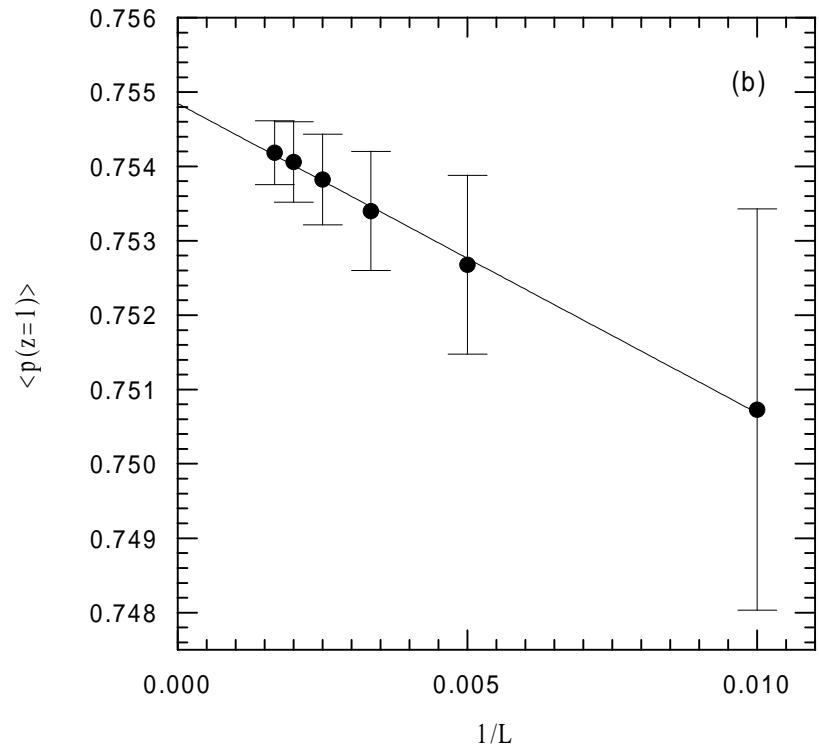
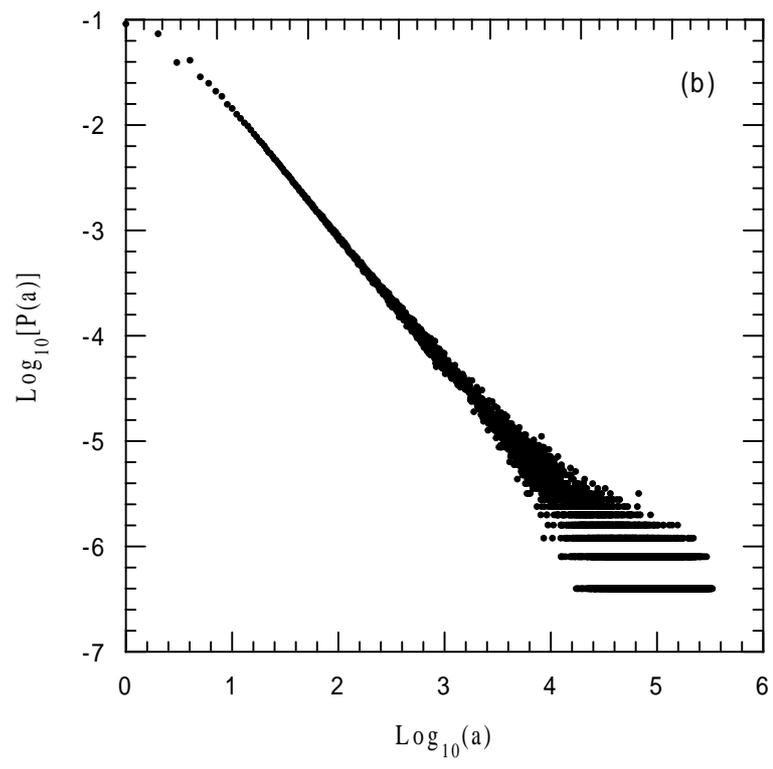
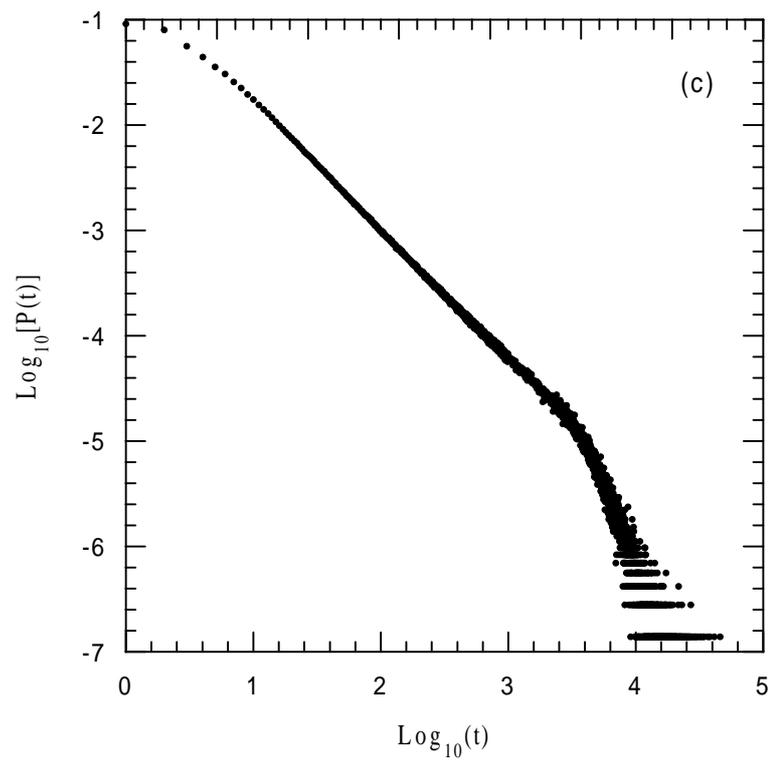
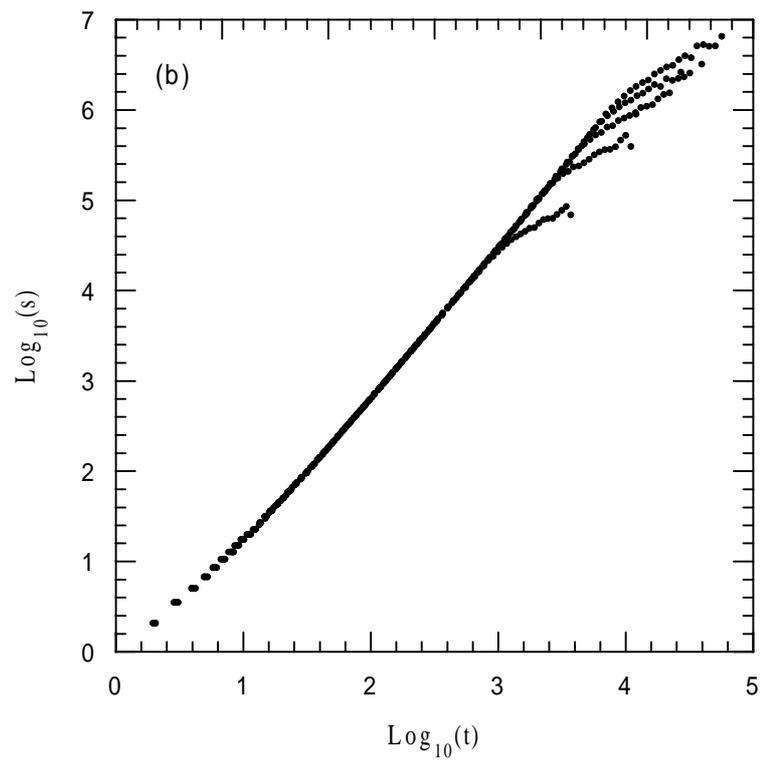
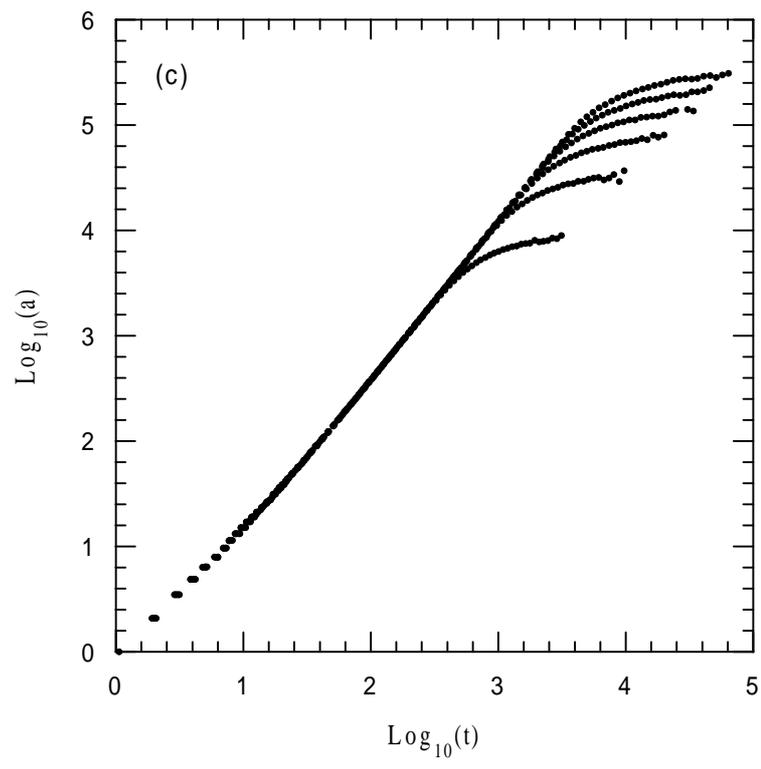
\end{document}